# Target Detection Framework for Lobster Eye X-Ray Telescopes with Machine Learning Algorithms

Peng Jia,[1,2] Wenbo Liu,[1] Yuan Liu,[3] and Haiwu Pan[3]

[1]*College of Physics and Optoelectronic, Taiyuan University of Technology, Taiyuan, 030024, China*
[2]*Peng Cheng Lab, Shenzhen, 518066, China*
[3]*National Astronomical Observatories, Beijing, 100101,China*

## ABSTRACT

Lobster eye telescopes are ideal monitors to detect X-ray transients, because they could observe celestial objects over a wide field of view in X-ray band. However, images obtained by lobster eye telescopes are modified by their unique point spread functions, making it hard to design a high efficiency target detection algorithm. In this paper, we integrate several machine learning algorithms to build a target detection framework for data obtained by lobster eye telescopes. Our framework would firstly generate two 2D images with different pixel scales according to positions of photons on the detector. Then an algorithm based on morphological operations and two neural networks would be used to detect candidates of celestial objects with different flux from these 2D images. At last, a random forest algorithm will be used to pick up final detection results from candidates obtained by previous steps. Tested with simulated data of the Wide-field X-ray Telescope onboard the Einstein Probe, our detection framework could achieve over 94% purity and over 90% completeness for targets with flux more than 3 mCrab ($9.6 \times 10^{-11}$ erg/cm$^2$/s) and more than 94% purity and moderate completeness for targets with lower flux at acceptable time cost. The framework proposed in this paper could be used as references for data processing methods developed for other lobster eye X-ray telescopes.

## 1. INTRODUCTION

Observations of transients in X-ray band could reveal some of the most violent activities in the Universe, which is an active research area for time domain astronomy. Telescopes of different configurations are proposed to detect photons in X-ray band, such as: coded aperture imager for the BAT on the Swift (Barthelmy et al. 2005) and the Eclair on the SVOM (Godet et al. 2014), the collimator for the Insight-HXMT (Zhang et al. 2020) and the pinhole camera for the MAXI (Matsuoka et al. 2009). In recent decades, the lobster eye telescope is proposed to carry out wide field sky surveys in X−ray band (Angel 1979). The lobster eye telescope is light-weighted and has very large field of view. With the development of micro−channel plate technology, lobster eye telescopes with micro−channel plates are proposed by many scientists for different space−based X−ray observation projects (Wilkins et al. 1989; Fraser et al. 1992; Kaaret et al. 1992; Chapman et al. 1993; Gehrels et al. 2011).

The MIXS was the first full lobster eye telescope used to observe celestial objects (Owens et al. 2001). Hereafter, many lobster eye telescopes are installed on pico or nano satellites (Van Inneman et al. 2000; Hudec 2018; Su et al. 2018) or the International Space Station (Tremsin & Siegmund 1998; Fraser et al. 2002) for X-ray observations. Thanks to these successful applications, lobster eye telescopes are now considered as important wide field X-ray survey instruments for many satellites, such as the MXT onboard the SVOM (Gotz et al. 2014), the WXT onboard the Einstein Probe (Yuan et al. 2018), the SXI onboard the SMILE (Peng et al. 2018), the LEXT onboard the Gamow

Corresponding author: Peng Jia
robinmartin20@gmail.com

(Feldman et al. 2021) and the SXI onboard the Theseus (Feldman et al. 2020). The point spread function (PSF) of lobster eye telescopes (Rhea et al. 2021) is very unique, which would spread photons to very large scale. With impacts brought by cosmic X-ray background, foreground diffuse radiation and limited photons from observation targets, it would be a challenge to design a proper target detection algorithm (Rees 1988).

Targets obtained by lobster eye telescopes will be normally observed by follow-up telescopes, which has high spatial resolution but small fields of view (Zhang et al. 2022). For example, the WXT (Wide-field X-ray Telescope) onboard the EP would carry out sky survey observations and the FXT (Follow-up X-ray Telescope) onboard the Einstein Probe (EP) would carry out the follow-up observations (Yuan et al. 2018). Therefore, two features are important for a proper target detection algorithm for data obtained by lobster eye telescopes. Firstly, the target detection algorithm should be robust to variations of observation conditions. Secondly, the target detection algorithm should have high purity (the purity indicates the percentage of real targets detected by the algorithm versus all detected targets) and moderate completeness (the completeness is the percentage of real targets detected by the algorithm to all real targets) to reduce the false trigger rate and thus improve the observation efficiency.

Traditionally, scientists would generate 2D images according to positions of photons on the detector plane and detect targets with classic algorithms. The SExtractor is a commonly used target detection algorithm (Bertin & Arnouts 1996). However, since the number of photons is small for celestial objects in X-ray band, the SExtractor would have relative low completeness in detection of celestial objects with low flux. Besides, the purity of the detection results would be also low, because there are many false targets generated by cruciforms of the PSF. To increase the purity and the completeness of the detection results, scientists would set a very low detection threshold for the SExtractor and design some hand made features to pick up detection results from candidates provided by the SExtractor. However, this method requires a lot of human interventions to select proper parameters and design proper features. Besides, the effectiveness of this method would be affected when data quality changes.

In recent years, scientists have proposed several new methods to process data obtained by lobster eye telescopes. Sawano et al. (2020) have proposed to project original data to 1D data and detect targets from 1D data. Results show that the method performs nearly 30% to 40% better than ordinary methods. Uchiyama et al. (2020) have proposed to detect and locate celestial objects from data obtained by multiplexing lobster eye optics, which could achieve almost 100% accuracy in determination of targets at 5 standard deviation detection limit flux (35 to 70 mCrab). Machine learning algorithms are another type of target detection algorithms, which have been widely discussed in recent years (Carleo et al. 2019). Because machine learning algorithms could detect targets after being trained with appropriate data, they would cost less human resources and are more robust to variations of data qualities. Different kinds of machine learning based target detection algorithms have been proposed and have achieved remarkable results (Gieseke et al. 2017; Gheller et al. 2018; Turpin et al. 2020; Ćiprijanović et al. 2021). Liu et al. (2022) have proposed and tested several different machine learning based target detection algorithms for data obtained by lobster telescopes. However results show that none of these algorithms could obtain satisfactory results.

In this paper, we propose a novel target detection framework. Firstly, several different machine learning based target detection algorithms will be used to detect celestial objects of different flux from 2D images. Then, the random forest algorithm will be used to obtain final detection results from detection candidates. We use 13193 simulated images of the WXT onboard the EP to test the performance of our framework (Yuan et al. 2018). Results show that our framework could achieve over 94% purity and moderate completeness for all targets and over 94% purity and over 90% completeness for targets with flux more than 3 mCrab ($9.6 \times 10^{-11}$ erg/cm$^2$/s). We will discuss details of our framework in this paper, which includes the following parts. In Section 2, we would describe basic properties of data obtained by lobster eye telescopes and the data pre-processing method. In Section 3, We will describe the target detection framework with multiple machine learning algorithms. In Section 4, We will show the performance of the detection framework. We will give our conclusions and anticipate our future work in Section 5.

## 2. DATA PROPERTIES AND THE DATA PRE-PROCESSING METHOD

We obtain the simulation data with the science simulator developed by Pan et al. (in preparation). The science simulator is a Monte-Carlo simulation code, which includes the photon and electron noise model and generates images through the ray-tracing method. PSF variations and responses of detectors are also considered in the simulator [1]. All X-ray targets are obtained from the ROSAT survey bright target catalogue (Voges et al. 1999). According to the design of the WXT onboard the EP, we could generate simulated data according to the pointing direction of the EP. The simulated data has four dimensions: arrival time, photon energy and coordinates of photons on the detector plane. The exposure time of simulated data ranges from 1100 seconds to 1300 seconds. The X-ray photons distribute in the detector plane, which has 4096 × 4096 pixels. The energy of these X-ray photons distributes from 0.5 to 4 kev. Although it would be possible to develop a framework to detect celestial objects of different types from simulated data (such as X-ray binaries, active galactic nuclei (AGNs) or Tidal Disruption Events (TDEs)), we would assume all celestial objects are the same in this paper and we could further develop classification algorithms with detection results. It should be noted that although we directly stack these data to 2D images for further process as shown in figure 1, it would be possible to design target detection algorithm directly for the 3D data, as we have discussed in Liu et al. (2022). The gray scale value of a pixel in 2D images is the number of photons that are detected by the detector in a particular direction. In figure 1, we can find that cruciform arms of targets with high flux could extend to very large scale. Targets with moderate or low flux would also distribute photons to large scale and therefore the grey scale values in pixels of these targets would be low.

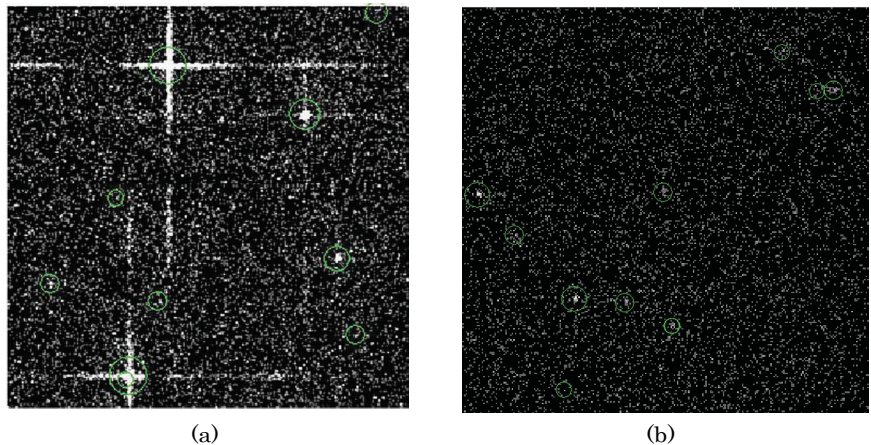

(a)                                           (b)

**Figure 1.** Figure a shows a two-dimensional stacked image containing a large number of targets with high flux and figure b shows a two-dimensional stacked image containing a large number of targets of moderate flux. The center of the green circle is the coordinate of these targets. Targets with moderate flux are hard to be seen, even we use the histogram equalization method to show these figures.

As we have discussed above, photons from a celestial object would distribute to a large scale in the detector plane. Therefore, images of celestial objects with moderate or low flux in original 2D images (4096 × 4096 pixels) would have many pixels with zero values. These images have low signal to noise ratio (SNR), which would be hard to be detected by target detection algorithms. According to our experience, we could bin the original image to a smaller size to increase SNR of celestial object images. Besides, since images of celestial objects with high flux would have very large cruciform arms, which would make it difficult to merge all these arms as one target. This problem could also be solved, if we resize original images to images of smaller size and detect targets from resized images. Therefore, we bin original images by 16 × 16 pixels and 8 × 8 pixels to generate images of 256 × 256 pixels (level-1 image) and 512 × 512 pixels (level-2 image) for the target detection framework. The efficiency of the detection algorithm could be increased, at the cost of lower positioning accuracy and lower detection efficiency for targets that are close to each other. It should be noted that since the size of window for binning depends on the flux of celestial objects, the pixel scale of binned images and the PSF of lobster telescopes. For data obtained with significantly longer exposure time, with PSF of smaller size, with larger pixel scale or with telescope of much larger collection area, we would set the binning window to a smaller size and vice versa.

## 3. THE STRUCTURE OF THE TARGET DETECTION FRAMEWORK

We have generated a level-1 image ($256 \times 256$ pixels) and a level-2 image ($512 \times 512$ pixels) from the same observation data for further target detection. For simulated data discussed above, these targets could be classified to three different types according to their brightness: ultra-bright targets, bright targets and ordinary targets. Ultra-bright targets are targets with flux more than 200 mCrab, bright targets are targets with flux more than 3.0 mCrab and ordinary targets are targets with flux more than 0.5 mCrab. In the original 2D images ($4096 \times 4096$ pixels), ultra-bright targets would distribute photons to more than $3000 \times 3000$ pixels, bright targets would distribute photons to more than $16 \times 16$ pixels and ordinary targets would distribute photons from $8 \times 8$ pixels to $16 \times 16$ pixels. If we have images with larger pixel scale or longer exposure time, we would reduce the binning window and use the same threshold to define targets of different flux. If we have images with PSFs of significant different shapes, we need to set a different threshold by hand. The level-1 image would be used for detection of ultra-bright and bright targets. The level-2 image would be used for detection of ordinary targets.

The diagram of the target detection framework is shown in figure 2. It includes four different parts:
1. The ultra-bright target detection part. Ultra-bright targets are targets with large cruciform arms, which could extend to the whole image. Because these targets are large and have obvious structure, we would use morphological operations to detect ultra-bright targets from level-1 images as candidates 1.
2. The bright target detection part. Bright targets are targets with cruciform arms, which could not extend to the whole image. We would mask candidates 1 with the PSF and use the Faster Region based Convolutional Neural Networks (Faster-RCNN) to detect bright targets from level-1 images as candidates 2.
3. The ordinary target detection part. Ordinary targets are targets that are not detected by previous steps. We would mask candidates 1 and candidates 2 with the PSF and use the Faster-RCNN to detect ordinary targets from level-2 images as candidates 3.
4. The target Classification part. We would apply weak restrictions to the detection results obtained by the Faster-RCNN and use the random forest (RF) to pick up true targets from all detected candidates (candidates 1, 2 and 3). With the framework, we could achieve high purity and acceptance completeness. We will discuss details of these steps in the following sub-sections.

### 3.1. *Performance Evaluation Criterion For Detection Results*

We use the purity and the completeness defined in equation 1 to evaluate the performance of different algorithms in our target detection framework.

$$Purity = \frac{TP}{TP + FP},$$
$$Completeness = \frac{TP}{TP + FN}. \tag{1}$$

We set the distance between prediction coordinates and label coordinates less than 2.0 (2.2 *arcmin*) pixels in the 2D image as the threshold for a true detection. If the distance is less than 2.0 pixels, we would set it as a true positive (TP) detection. Otherwise, it would be either a false positive (FP) or false negative (FN) detection. We set the prediction result is positive for positive sample as the True Positive (TP), the prediction result is positive for negative sample as False Positive (FP), the prediction result is negative for negative sample as False Positive (TN) and the prediction result is negative for positive sample as False Negative (FN). It should be noted that our detection algorithm framework is used to obtain rough position of celestial objects and we could further obtain positions of celestial objects with high accuracy through regression algorithms (Jia et al. 2021).

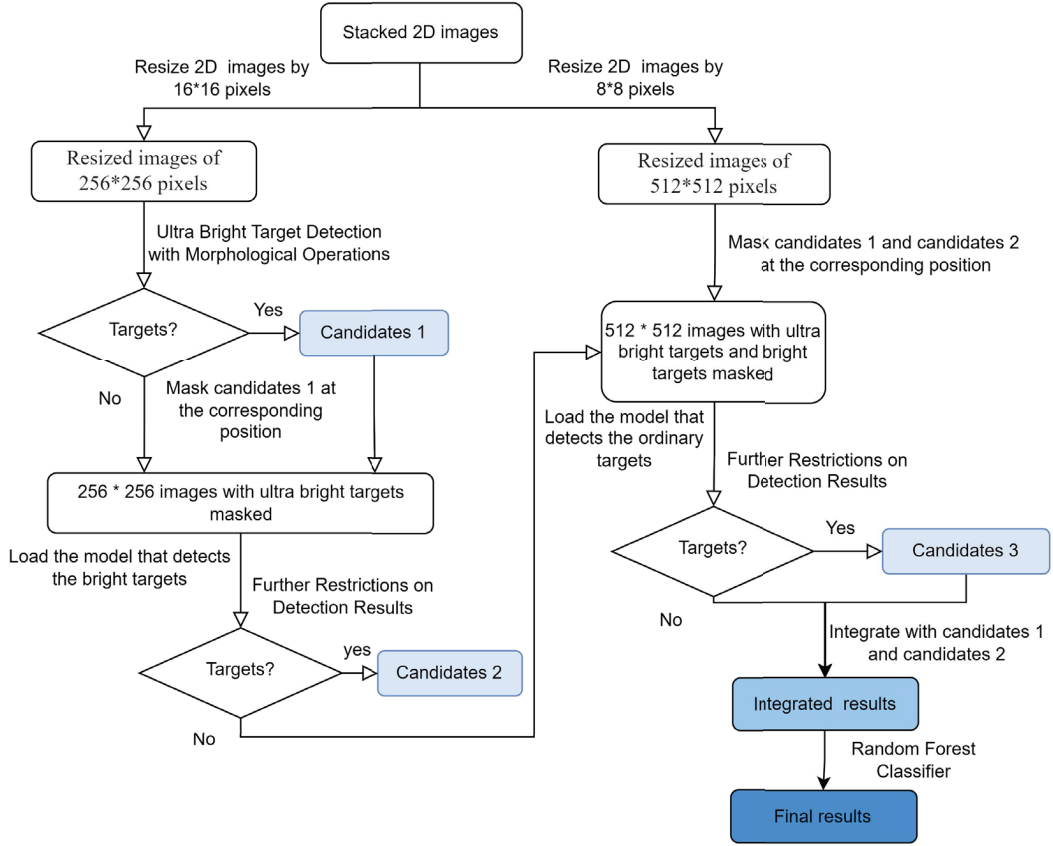

**Figure 2.** The diagram of the target detection framework. The whole process could be divided into four steps and two images with different pixel scales are used for target detection.

### 3.2. *Ultra Bright Target Detection with Morphological Operations*

Ultra bright targets would spread photons to very large scale and there would be many detection alarms for the same target, as shown in figure 3. It would be quite difficult to merge all detection alarms as one target. Therefore, we would use morphological operations from the OpenCV package to detect ultra bright targets from level-1 images (Itseez 2015), which includes the following steps:

1. Erosion. We would carry out the erosion operation on level-1 images with kernel size of $3 \times 3$ pixels, which would set gray scale of all pixels within the $3 \times 3$ pixels as the minimal gray scale value within the kernel. The erosion step would remove most of the isolated noise points. Because this step is majorly used to remove point like noises, the kernel size is independent of observation conditions or PSFs. Therefore, parameters are the same in this step, if we process data from different lobster eye telescopes.

2. Filtering. We would then convolve eroded images with a 2D Gaussian kernel with size of $3 \times 3$ pixels to enlarge images of ultra-bright targets. In this step, the size of convolutional kernel is also independent to observation conditions or PSFs and we could set it as a constant number to process data from different lobster telescopes.

3. Binarization. An appropriate threshold is set for the filtered image to obtain a binary image. Pixels with gray scale values that are less than the threshold will be directly set to 0, while pixels with gray scale values that are larger than the threshold will be set to 1. Pixels in the binary image with gray scale values that are not zero belong to images of celestial objects. The threshold is directly related to the flux of observed targets, the PSF and the pixel scale of the lobster eye telescope. Normally we would expect to obtain the central region of celestial objects images, which includes the central focus and small part of the cruciform arms. Therefore, we would firstly obtain the maximum gray scale value of ultra-bright targets whose flux are about or more than 200 mCrab. Then we would calculate average value of these maximal values as the binarization threshold.

4. Dilation. Dilation is the inverse operation of the erosion, which would set all pixels in the image covered by a

kernel as the maximal value within the image. We would dilate binary images with a Gaussian kernel, which could merge images of the same target as one target. The kernel size is related to the size of the gap and the width of the cruciform arm. In real applications, we would set the size of kernel a little larger than the width of the cruciform arm or that of the gap.

5. Contour detection. At last, we would use the contour detection algorithm from the OpenCV to obtain contour points of ultra-bright targets. The contour point with the largest grey scale value that is surrounded by other contour points will be set as location of targets.

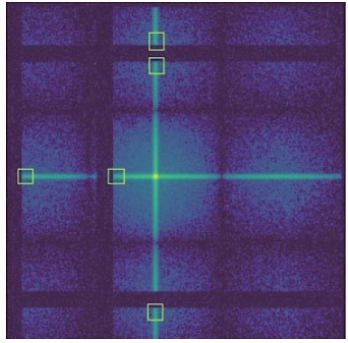

**Figure 3.** This figure shows a stacked image with an ultra bright target. Cruciform arms of the bright target would be affected by gaps between micro channel plates (labeled by yellow boxes), which would introduce false detection results.

For the ultra bright target detection step, the threshold in the binarization algorithm and the size of kernel in the dilation algorithm need to be specifically adjusted when the observation telescope is different. For simulated data of the WXT onboard the EP with exposure time of around 1200 seconds, according to the above binarization threshold calculation method, we set the binarization threshold to 300, and set the kernel size in the dilation step as 21 × 21 pixels. With parameters defined above, we could achieve 100% purity and 100% completeness in 48 test images which have 11 ultra-bright targets. One of these images being processed by different steps is shown in figure 4. As shown in this figure, false detection results generated by arms from different celestial objects and noise could be effectively removed with the first two steps. Then with the binarization and dilation step, we could obtain positions of ultra bright targets. Besides, we could find that some targets of moderate or low flux are removed in this step. However, since we would require to obtain results with high purity and moderate completeness in this part and we could use methods discussed later to detect these targets, it is acceptable to remove them.

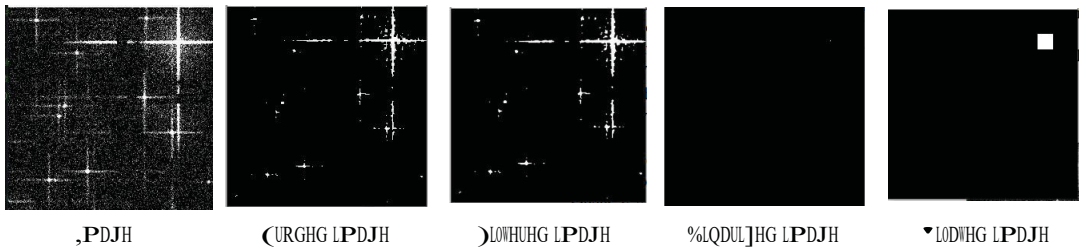

**Figure 4.** The position of the ultra bright target could be obtained after a series of steps discussed above.

We further use a total of 13193 images to test the morphological operation algorithm and we have obtained 98% completeness and 82% purity. We have checked all false detection results. We find that bright targets close to each other or targets with low positioning accuracy would lead to low completeness. Meanwhile, cross parts between arms from different targets would lead to false positive detection results, however these targets would be removed when we use the RF to pick up true targets from all detection results.

### 3.3. *Bright and Ordinary Targets Detection with the Faster-RCNN*

#### 3.3.1. *Introduction to the Faster-RCNN*

The Faster-RCNN was firstly proposed by He et al. (2016) for general purpose target detection tasks and has been widely studied and applied in different applications. Since it has better performance in detection of small targets, Jia et al. (2020) has proposed a modified Faster-RCNN based algorithm to detect celestial objects from images obtained by wide field optical telescopes. The Faster-RCNN has also been proposed to detect targets from data obtained by radio telescopes (Lao et al. 2021). These successful applications show the advantage of the Faster-RCNN algorithm in detection of small targets. In this paper, we would also use the Faster-RCNN as the basic block to detect bright and ordinary targets. Since the detection threshold could be modified accordingly, we would set a smaller detection threshold to obtain candidates of celestial objects with high completeness and low purity, and then we would pick up true targets from these candidates with other algorithms in the following step.

The structure of the Faster-RCNN used in this paper is shown in figure 5. It includes three parts: feature extractor (Extractor), Region Proposal Net (RPN) and Region of Interest Alignment Part (ROI Align). Images would be sliced to small parts and features of each part of these images would be extracted by the Extractor and sent to the RPN. ResNet50 (Residual Network with 50 layers) is used as the Extractor in this paper (He et al. 2016). The ResNet50 uses shortcut connection to keep the neural network easier to train even with many layers. We modify the convolution kernel size in Layer0 from 7 × 7 to 3 × 3 to increase its ability in detection of small targets. The modified ResNet50 is shown in figure 6. In this paper, we adapt the feature pyramid network to further increase detection efficiency (Lin et al. 2017), which would integrate multi-scale features for further detection. We set all normalization layer as Group Batch Normalization and Leaky Rectified Linear Units (Leaky ReLU) function as activation function (Maas et al. 2013) in the Faster-RCNN. The advantageous of the Leaky ReLU is that when the input value is negative, the back propagation of the neural network can still obtain a small gradient so that the parameters can be further updated.

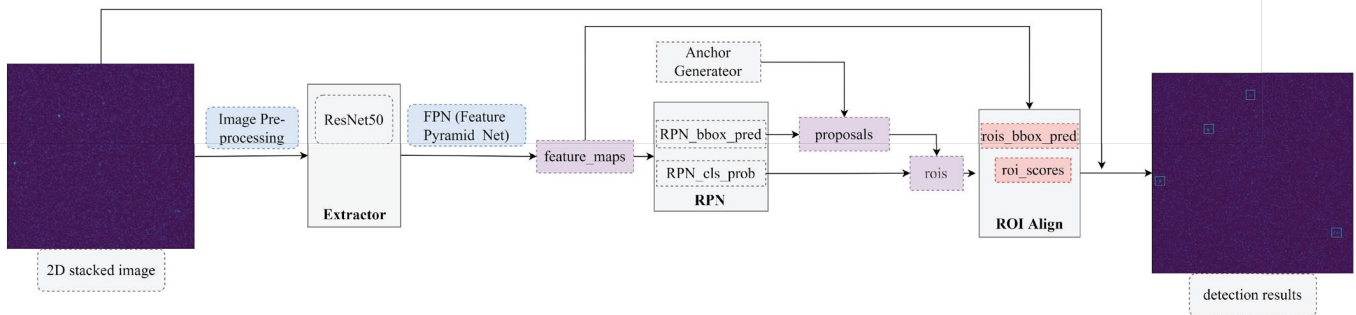

**Figure 5.** The structure of the Faster-RCNN used in this paper. It includes three parts: feature extractor (Extractor), Region Proposal Net (RPN) and Region of Interest Alignment part (ROI Align).

In the Faster-RCNN, the data will be sent to the RPN after being processed by the Extractor. There are two branches in the RPN as shown in figure 7. The RPN_cls_pred uses softmax classifier to classify detection box either to foreground or background. The RPN_bbox_pred calculates appropriate shift for each detection box to obtain higher scores. The anchor generator would generate several anchors at different positions for further classification. Then we would pick up proposals of foreground or targets with top-n possibilities. Finally we would use the non-maximum suppression algorithm (NMS) (Neubeck & Van Gool 2006) to get the region of interest (rois) from these proposals. With the NMS, we could increase detection accuracy and speed, but if some targets are near a brighter target, the detection results will be suppressed. However, it is a trade-off that we must make. Otherwise, there would be huge amount of detection results which would excess the maximal number of candidates a Faster-RCNN could output.

The RPN part would output rois of different size. It would be hard to design classification and regression algorithm for rois with different size. The roi alignment part would set feature maps of rois to the same size with bi-linear interpolation (He et al. 2017) as shown in figure 8. Then, we would carry out further classification and position regres-sion with resized rois to get final results. In this paper, we would only classify candidates as background or targets. With steps mentioned above, we could obtain targets from observation images. Since images of celestial objects with different flux would have different size, particularly for the cruciform arms, we would use the Faster-RCNN to detect bright and ordinary targets in a multiple scale way. In each step, detected targets (candidates) will be masked with the PSF and we will detect targets from masked images.

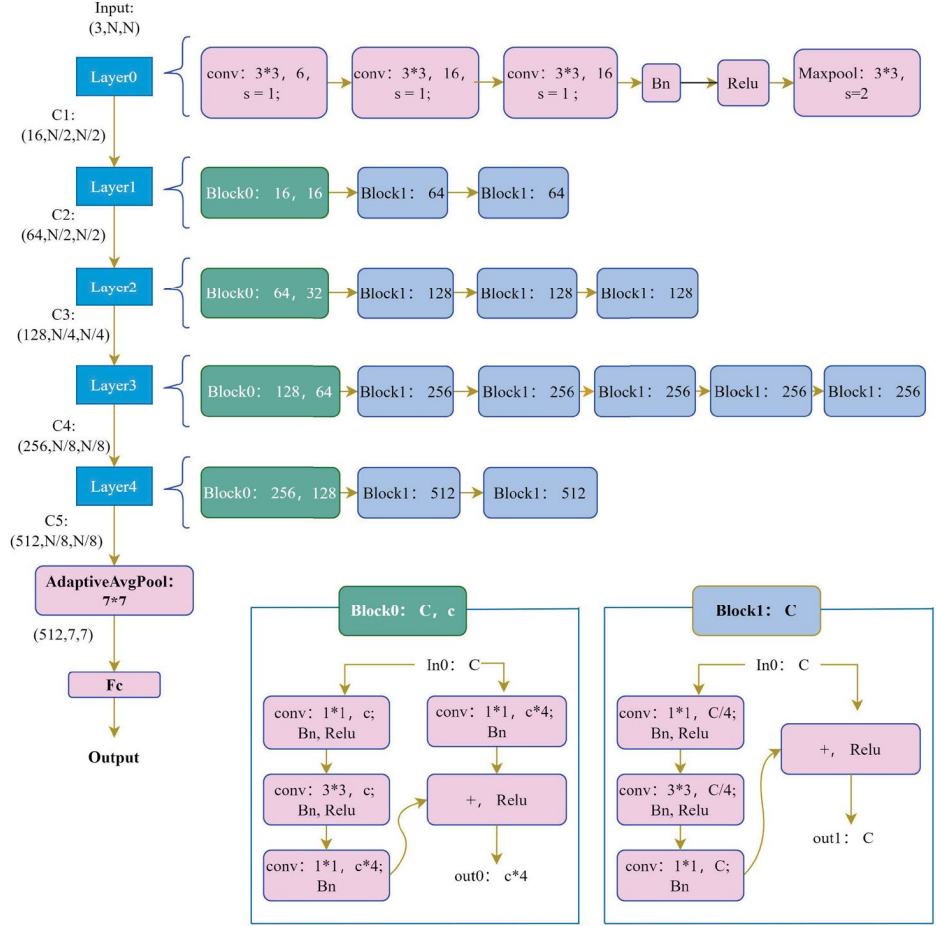

**Figure 6.** The structure of the modified ResNet50 used in this paper. The ResNet50 is used as feature extraction neural network. The structure of different layers is shown as blocks with size, channel, normalization method and activation function. In the figure, "conv" represents the convolution layer, and the numbers represent the size and number of the kernel respectively, "fc" represents the fully connected layer, and "Bn" represents Group Batch Normalization, "Relu" represents Leaky ReLU, "s" is stride, which represents the moving step. Block0 and Block1 represent two different connection blocks, the Block0 block has two input parameters and the Block1 has one input parameter.

### 3.3.2. *Bright Target Detection with the Faster-RCNN Algorithm*

In this step, we would detect bright targets as candidates 2 from level-1 images (256 × 256 pixels). First of all, we would generate a mask matrix with the same size as that of the level 1 image, except that we would place simulated stamp images in places of candidates 1 in the mask matrix. We generate stamp images through multiplying the mask threshold with the PSF. Then we would compare gray scale values of the level 1 image with those in corresponding mask matrix and set gray scale values of the level 1 image as zero, if gray scale values of each elements in the mask matrix are equal to or smaller than gray scale values of the level 1 image. For data obtained by different telescopes, the mask threshold could be modified from 0.2 to 1. With higher threshold, we would mask more pixels and vice versa. For simulated data of the WXT on board the EP, 0.5 is the best value. A masked image is shown in figure 9.

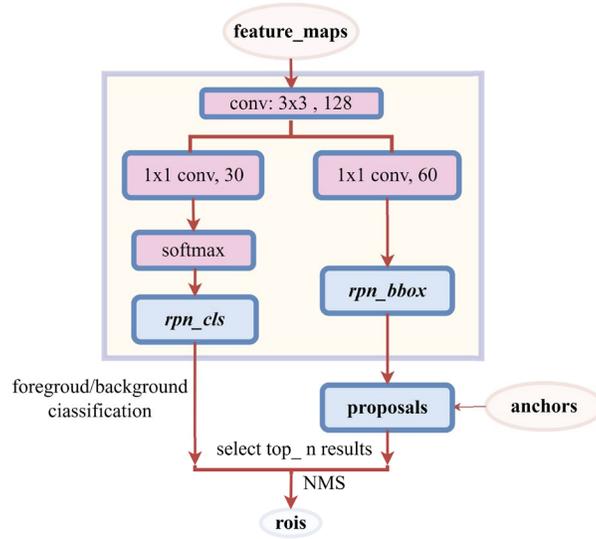

**Figure 7.** The structure of RPN. The left branch is used to propose regions which may include targets. The right branch is used to calculate appropriate positions of proposed regions. Finally we choose the suitable rois from them. In the figure, "feature_maps" represents a set of multi-scale feature layers, "conv " represents the convolution layer and the numbers represent the size and depth of the convolution kernel in turn, "rpn_cls" represents the two categories predicted by the rpn layer, "rpn_bbox" represents the two diagonal coordinates of the bounding-box predicted by the rpn layer.

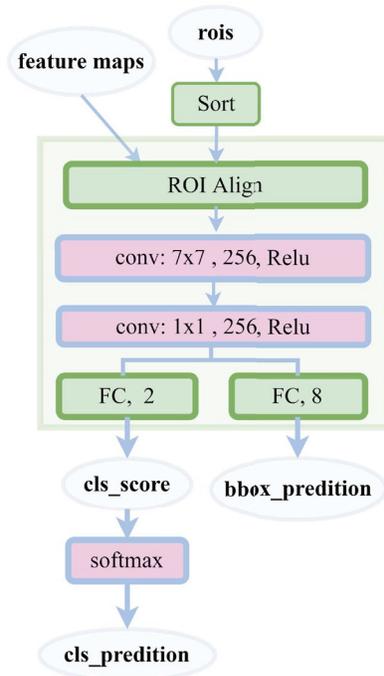

**Figure 8.** The structure of roi Alignment. Feature maps of different spatial scales and region of interests (rois) will be sent to the roi alignment part. In the roi alignment part, we would carry out classification (the left branch) and bounding-box regression (the right branch) to obtain final detection results. In the figure, "feature maps" represents a set of multi-scale feature layers, "Sort" represents sorting the scores of rois from large to small, "conv " represents the convolution layer and the numbers represent the size and depth of the convolution kernel in turn, "FC" represents the fully connected layer and the number represents the number of output nodes in the full connection layer, and "Relu" represents Leaky ReLU.

**Table 1.** Purity and completeness of the Faster-RCNN in bright targets detection with different confidence.

| Confidence threshold | Flux interval | Completeness | Purity |
|---|---|---|---|
| 0.88 | More than 1(0-1) | 0.8532(0.1558) | 0.90 |
| 0.90 | More than 1(0-1) | 0.8455(0.1327) | 0.94 |
| 0.92 | More than 1(0-1) | 0.8312(0.1245) | 0.96 |

We can find that bright targets are more easily to be detected after we mask ultra bright targets.

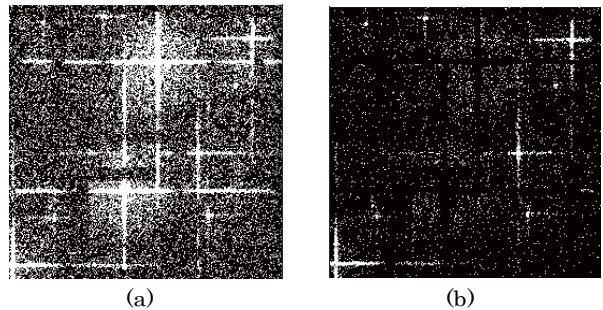

(a)                                        (b)

**Figure 9.** Figure a shows the original level-1 image and figure b shows the same image after we mask ultra-bright targets. As can be seen from the figure, bright targets are easier to be detected after we mask ultra-bright targets.

We use the Faster-RCNN discussed in section 3.3.1 for bright target detection. Since in this step, we focus on detection of bright targets, we would mask ultra-bright targets and only label bright targets as labels in the training data set. We use 480 images as the training data and train the neural network by 46 epochs with Adam Optimizer as its optimizer (Kingma & Ba 2014), which will take around 1945.8 seconds in a computer with a GTX 1080 Ti. After training, we use 384 images to test the performance of the Faster-RCNN in detection of bright targets. The confidence is the detection threshold used in the Faster-RCNN to make a trade-off between completeness and purity. We have tested the performance of the Faster-RCNN with different confidence. With different confidence, the corresponding purity and the completeness for targets with different flux are shown in table 1. It can be seen that when the confidence threshold is 0.92, the purity of the Faster-RCNN is 0.96 and the completeness of the Faster-RCNN in detection of targets with flux above 1.0 mCrab can reach more than 0.83, which is acceptable for us. For data obtained by other telescopes, we need to plot the relation of the purity and the completeness between the threshold and set an appropriate value accordingly.

### 3.3.3. *Ordinary Target Detection with the Faster-RCNN Algorithm*

After we have detected ultra bright and bright targets with previous steps, we would further use the Faster-RCNN to detect ordinary targets as candidates 3 from level-2 images (images of $512 \times 512$ pixels). We would firstly mask ultra-bright and bright targets in these images with the method discussed above as the training data. Then we would train another Faster-RCNN neural network with these masked images. One image before and after masking are shown in figure 10. We train the Faster-RCNN with 480 images by 13 epochs with Adam Optimizer as its optimizer (Kingma & Ba 2014), which will take around 549.9 seconds in a computer with one GTX 1080 Ti.

We use 384 images to test the performance of the Faster-RCNN in detection of ordinary targets with different confidence thresholds. The purity and the completeness in detection of ordinary celestial objects are shown in table 2. It can be seen that when the confidence threshold is 0.95, the purity of the whole ordinary target detection algorithm is 0.84. At the same time, the detection completeness of targets with flux above 1.0 mCrab can reach more than 0.56, which is acceptable for us. Targets with flux lower than 1.0 mCrab would be too dim to be detected and only in rare cases, when photons are gathered in the centre, the Faster-RCNN could detect these targets.

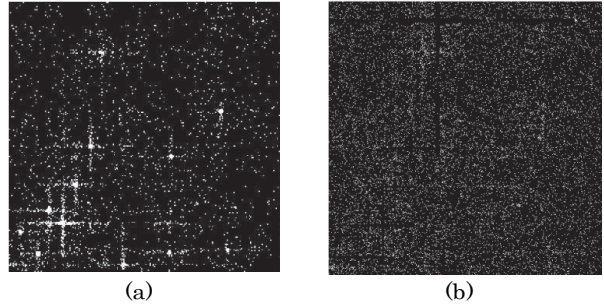

(a)                                                      (b)

**Figure 10.** Figure a shows the level-2 image and figure b shows the level-2 image after we mask bright and ultra-bright targets. As can be seen from the figure, ultra bright and bright targets are masked with our method.

**Table 2.** Purity and completeness of the Faster-RCNN in detection of ordinary targets with different confidence.

| Confidence threshold | Flux interval | Completeness | Purity |
|---|---|---|---|
| 0.88 | 0-1(More than 1) | 0.0802(0.5894) | 0.66 |
| 0.92 | 0-1(More than 1) | 0.0690(0.5856) | 0.75 |
| 0.95 | 0-1(More than 1) | 0.0447(0.5689) | 0.84 |

### 3.3.4.  *Sensitivity Analysis for Parameters in the Faster-RCNN Algorithm*

The Faster-RCNN is a supervised learning algorithm. Comparing with other target detection algorithms, the Faster-RCNN could learn features directly from observation data for target detection, which makes it robust in real applications. However, there is still one parameter (the confidence) in the Faster-RCNN, which needs to be carefully selected to assure their best performance. To test the sensitivity of the confidence in the Faster-RCNN, we fix all parameters of one of the Faster-RCNN models, and then adjust the confidence of the other Faster-RCNN. The confidence of the bright target detection model ranges from 0.895 to 0.945 (the confidence of the ordinary target detection model is a fixed value of 0.95). The confidence of the ordinary target detection model ranges from 0.925 to 0.975 (the confidence of the bright target detection model is a fixed value of 0.92). We adjust the confidence with a step of 0.05 and obtain corresponding purity and completeness of these two models in figure 11.

From this figure, we can find that the confidence of the ordinary target detection model has little influence on the detection results, while the whole detection purity and completeness metrics are more sensitive to the confidence of the bright target detection model. As this confidence increases, both purity and completeness will change significantly. Therefore, we need to carefully set the confidence of the bright target detection model according to real requirements.

### 3.4.  *Further Weak Restriction Conditions on Detection Results from the Faster-RCNN*

We could further design several restriction conditions based on physical prior conditions (such as the morphology of the PSF or the pixel scale) to further select true targets from candidates obtained by previous steps. We would extract stamp images of detected targets from the original image and resize these images to 5 × 5 pixels. Then, as shown in figure 12, we would further select candidates according to the following conditions:

1.The photon distribution condition. We would calculate the total number of photons in the central column and the central row of extracted images and the total number of photons in extracted images. We would then calculate divisions between these two values and set the threshold of 0.38 for bright targets and 0.2 for ordinary targets to classify between candidates and false alarms. The photon distribution condition is directly related to the PSF and the pixel scale and we need to modify this condition according to data properties from different lobster eye telescopes.

2.The correlation condition. With candidate images, we would calculate correlations between sub-images and the specified PSF. We would set the threshold of 0.6 for bright targets and 0.1 for ordinary targets to classify between candidates and false alarms. The PSF of lobster eye telescope is directly related to this condition and we need to use different PSFs for images obtained by different lobster eye telescopes.

3.The photometric condition. For an object on a single image, we would use circular aperture photometry with radius of $R_1$ (aperture area of $S_1$) to obtain flux $N_1$ and we would also use circular aperture photometry with radius of $R_2$ (aperture area of $S_2$) to obtain flux $N_2$. Then we will calculate NetFlux defined in equation 4 and set the threshold of 8 as the classification threshold between candidates and false alarms. The photometric condition is also directly related to the PSF of lobster eye telescopes, we need to set different conditions for data obtained by different telescopes.

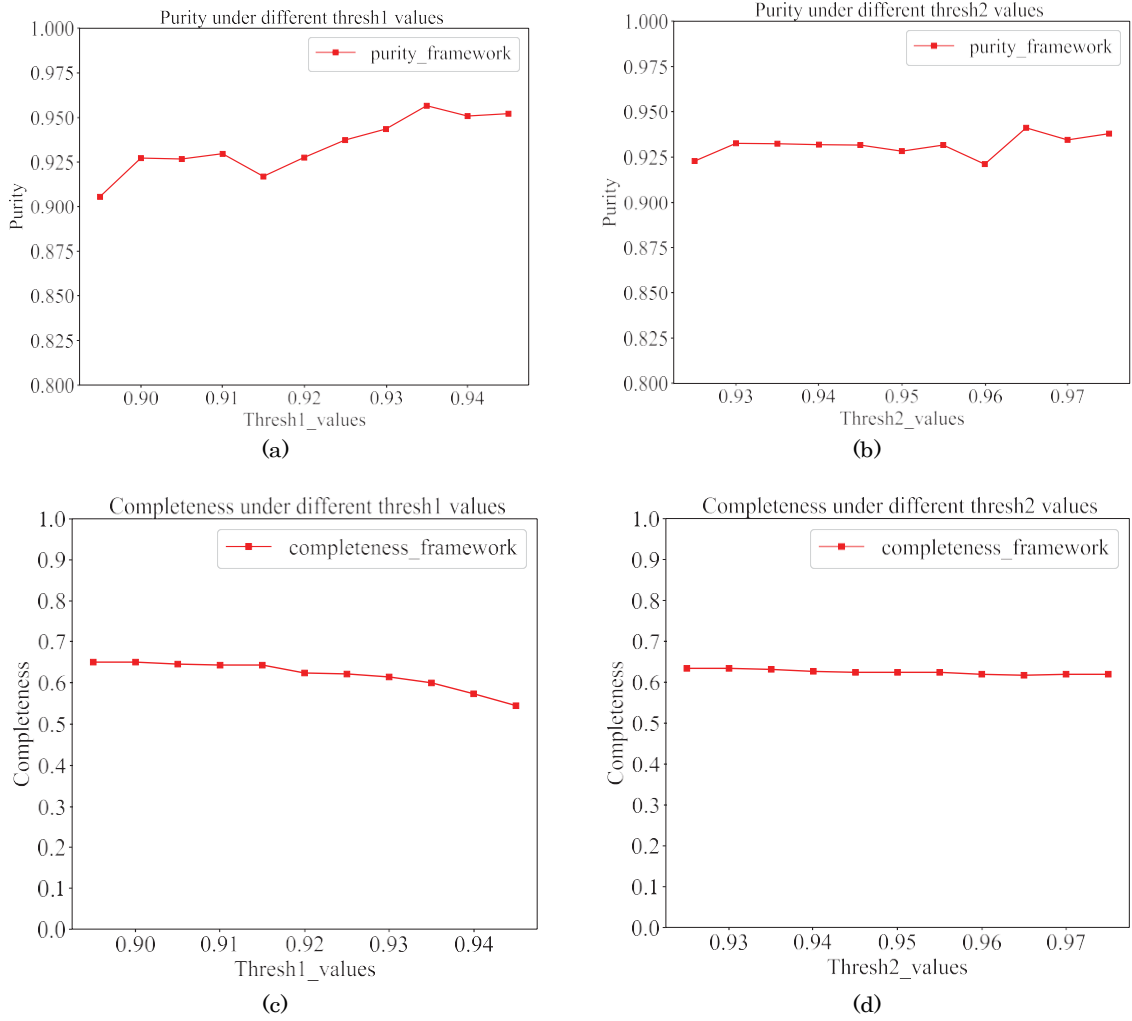

**Figure 11.** The purity and the completeness of two Faster-RCNN detection models with different confidences. Figure a and figure c show the purity and the completeness of our framework in detection of targets with different confidence values in the Faster-RCNN for bright target detection. Figure b and figure d show the purity and the completeness of our framework in detection of targets with different confidence values in the Faster-RCNN for ordinary target detection. We can find that the confidence of the bright target detection model is more sensitive than the ordinary target detection model.

$$NetFlux = N_1 - S_1 \cdot \frac{N_2}{S_2}. \tag{2}$$

These restriction conditions are directly related to specific telescopes that are used to carry out observations. Several experiments need to carry out to select proper parameters. However, these parameters are relatively robust for the same telescope with different PSFs, according to our experiments.

### 3.5. *Selection of Detection Results with the Random Forest Algorithm*

Restriction conditions discussed in Section 3.4 are only weak conditions, which will remove false detection results that are obviously different from true targets. However, there are still some false detection results similar to true

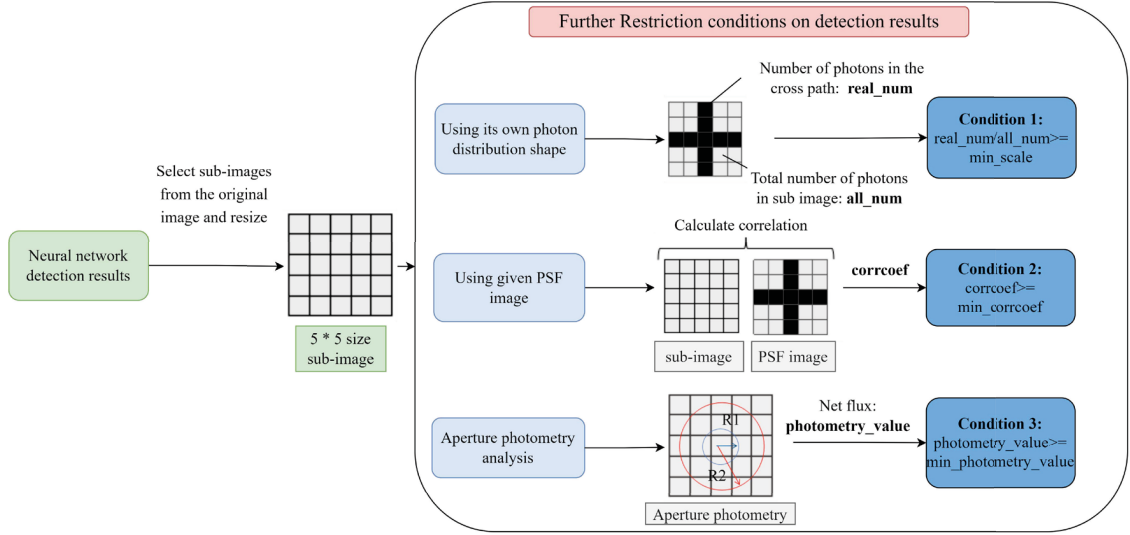

**Figure 12.** The schematic diagram of further weak restriction conditions, which includes the photon distribution condition, the correlation condition and the photometric condition

targets. Therefore, we use the Random Forest algorithm (RF) to remove false alarms that are similar to real target to further increase the purity of the detection framework. The RF is a supervised machine learning algorithm, which builds many independent decision trees by random sampling (breiman L. 2001). The key to distinguishing the RF from the simple average of all decision trees lies in two concepts: random sampling of training set samples and selection of random feature subsets when dividing nodes. The RF can reduce the risk of over-fitting and we could use parallel computation to increase the time efficiency of the algorithm. The RF is widely used as an effective classification algorithm in many applications (Gao et al. 2009; Carrasco et al. 2015; De Cicco et al. 2021).

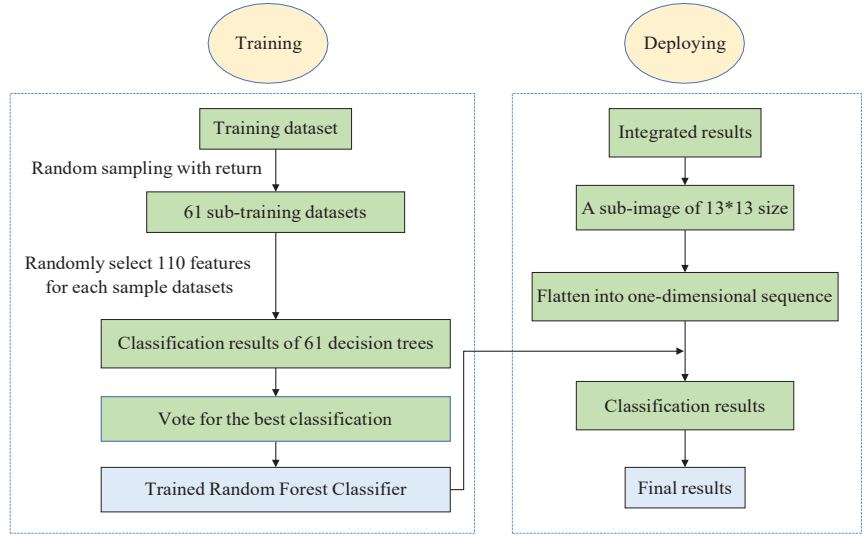

**Figure 13.** The schematic draw of training and deploying of the RF. We set the number of decision trees as 61 and the number of features is defined automatically by the random forest algorithm in the Scikit−learn package.

The process of training and deploying the RF is shown in figure 13. The training process includes the following steps:

1. Images of all candidates will be cropped from level−2 images by $13 \times 13$ pixels (the value of 13 is the median size

of these targets). Then each sub-image would be flattened into an one-dimensional sequence with 169 variables.

2. We randomly select 6678 training samples and use bootstrap sampling method to select the training set of a decision tree, and perform 61 rounds of sampling process.

3. We randomly select 110 features from 169 features in the feature space to form a new feature set and train 61 weak classifiers in parallel with new feature set.

4. Since each weak classifier is independent of each other, we will build the final classification algorithm with votes from all the decision trees.

We use the RF defined in the Scikit–learn package (Pedrogosa et al. 2011) and use a data set with 5010 true targets and 1668 false targets to train the RF, which would take 4.08 seconds. We use the Receiver Operating Characteristic curve (ROC curve) and Area Under roc Curve (AUC) as the criteria to measure the performance of the classification model (Mandrekar 2010). The Abscissa of the ROC curve plane is false positive rate (FPR), and the ordinate is true positive rate (TPR). The definitions of FPR and TPR are shown in equation 8:

$$FPN = \frac{FP}{N}$$
$$TPR = \frac{TP}{P}. \qquad (3)$$

Where $P$ is the total number of real positive samples and $N$ is the total number of real negative samples. We further plot the ROC curve of the RF as shown in figure 14. The value of AUC is the size of the area below the ROC curve, and the closer the AUC is to 1, the better the performance of the classifier. As can be seen from the figure, our classifier has an AUC of 0.92 on the test set, which is acceptable for our applications. At the same time, it can be seen that the ROC curve has an obvious elbow where we could select a threshold with balanced performance in both the TPR and the FPR. We use 0.55 as the threshold for the RF, and the confusion matrix on the test set is shown in figure 15. The confusion matrix consists of two dimensions: the actual category of the sample and the category predicted by the classifier model (Visa et al. 2011). We use the accuracy defined in equation 4 to evaluate the performance of the algorithm on the test set. The accuracy of the RF in the test set could be around 91%, which satisfies our requirement.

$$accuracy = \frac{TP + FN}{TP + FN + TN + FP}. \qquad (4)$$

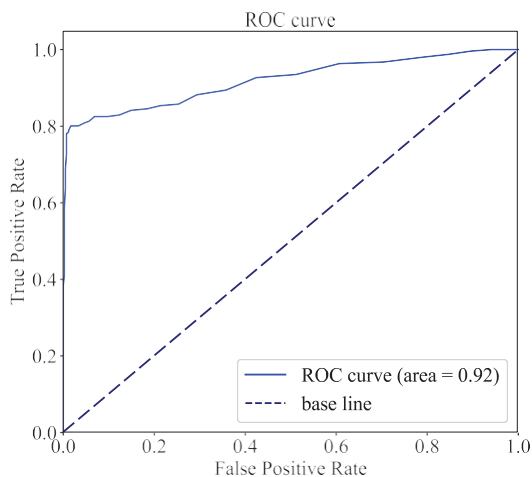

**Figure 14.** ROC curve of the random forest classifier on the test set.

In order to investigate the importance of features in the RF, we use variable importance measures (VIM) as an evaluation metric. The larger the VIM, the more important the classification function of the feature in the RF model. The VIM is usually measured by the Gini index (the smaller the Gini index, the purer the result of a node segmentation according to a certain feature). The importance of a feature variable $X^j$ at node q of the tree i is expressed by the change of Gini index (GI) before and after the node q branch as defined in equation 8,

$$VIM_j^i = GI_q^i - GI_l^i - GI_r^i \tag{5}$$

where $GI_q$ represents the Gini index at node q, and $GI_l$ and $GI_r$ represent the Gini index of two new nodes branched by node q, respectively. The Gini index of node q of the tree i could be caluclated with equation 6,

$$GI_q^i = \sum_{c=1}^{C} \sum_{c' \neq c}^{C} p_q^i \cdot p_{q'}^i = 1 - \sum_{c=1}^{C} (p_q^i)^2, \tag{6}$$

where C indicates the number of categories and $q_q^i$ represents the proportion of category c in node q. In this paper, we set C as 2. If the set of nodes in which the variable $X^j$ appears in decision tree i is Q, then the importance of $X^j$ in the whole random forest algorithm is:

$$VIM_j = \sum_{i=}^{k} VIM_j^i = \sum_{i=1}^{k} \sum_{q \in Q} VIM_{i_j}^i, \tag{7}$$

where k is the number of decision trees in RF and k is 61 in this paper. Finally, importance scores of all features are normalized and sorted from large to small. We select features whose VIM is greater than the average VIM and we can find that features locate at the edge and the center of the image of celestial objects play important roles in the RF.

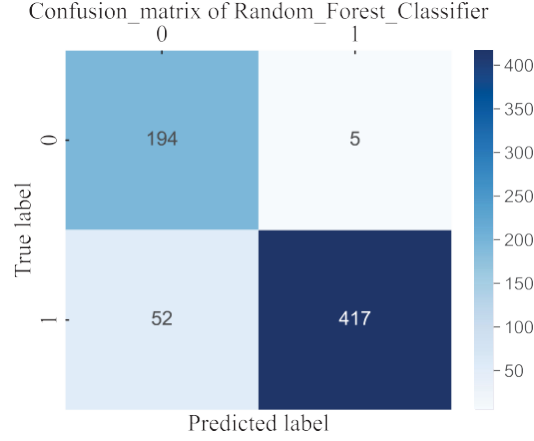

**Figure 15.** Confusion matrix for the random forest classifier on test set.

## 4. PERFORMANCE TEST OF THE DETECTION FRAMEWORK

In this section, we would use a new set of simulated data of the WXT onboard the EP and use the whole framework to detect targets from simulated data to test the performance of the framework. All machine learning algorithms in the framework are trained with another set of simulated data with the same configuration. We would set a flag in the position of a candidate to label candidates detected by different algorithms in the framework. In this circumstance, there is a little chance that some targets that are close to each other would be set as a correct detection which would reduce the completeness. Meanwhile, some targets would be detected several times, which would reduce the purity (since these detection results would be far from the center). However, as long as the position of one target is obtained, other targets can be further identified in subsequent observations.

We have generated 13193 images, with exposure time of around 1200 seconds, to test the performance of our algorithm. There are 20639 targets (with flux more than 0.5 mCrab) in these images. The detection steps and the purity in these steps are shown in figure 16. With the ultra-bright target detection step, we obtain 363 TPs and 78 FPs. Then we use the first Faster-RCNN to detect bright targets and obtain 6475 TPs and 517 FPs. Meanwhile, we detect ordinary targets with the second Faster-RCNN and obtain 6419 TPs and 1903 FPs. We use the restriction conditions to remove 27813 false alarms and 3849 true targets. At last, we use the random forest to remove 1873 candidates and 94 of them are true targets. In total, our framework could detect a total of 13617 targets, of which 12880 are true targets, with a purity of 94.59% and a completeness of 62.41%.

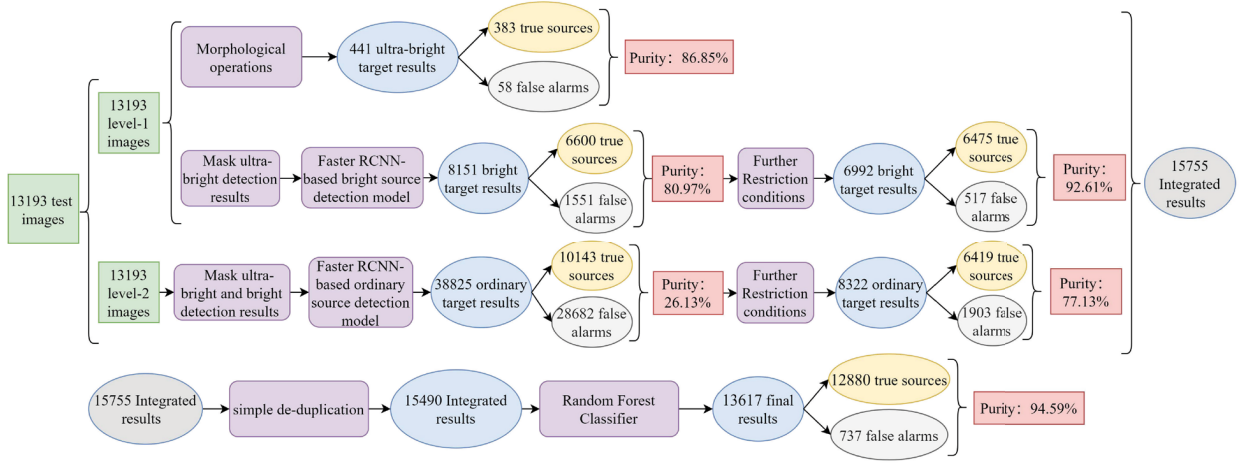

**Figure 16.** Schematic draw of the detection framework and the purity in different steps.

For comparison, we use the SExtractor based method to detect targets from simulated images. The detection process of the SExtractor includes background estimation, noise calculation, background deduction, filtering, source deblending and locating (Bertin & Arnouts 1996). Since the SExtractor is designed to process images obtained by optical telescopes, we have modified the SExtractor accordingly to achieve a completeness of 26.96% (70% for targets with flux of more than 3.0 mCrab) and a purity of 30.81%. Since the purity is relatively low, we further use a specially designed RF to pick up true targets from candidates obtained by the SExtractor. Overall, the SExtractor method could find a total of 6078 targets with 5464 real targets (purity of 89.50% and completeness of 26.47%).

We further plot the purity-completeness curve of our detection framework and that of the SExtractor for targets with different flux ranges in figure 17. The curves show that our detection framework has better performance in detection celestial objects in different flux ranges, especially ultra bright sources. Since detection results obtained by lobster eye telescopes are majorly used for transient observation, the purity is more important than the completeness. To better show the performance of our framework, we use the F0.5 score defined in equation 8 as the evaluation criterion.

$$F0.5 = \frac{5 \cdot purity \cdot completeness}{purity + 4 \cdot completeness}. \tag{8}$$

We plot the curve of the completeness and the F0.5 score for our detection framework and the SExtractor-based detection algorithm with different flux in figure 18. We could find that the performance of our framework is also obviously better than that of SExtractor-based detection method for celestial objects with flux close to the observation limit of lobster eye telescopes (i.e., flux more than 0.5 mCrab and below 1.0 mCrab). We can also find that the role of the RF based classifier is different for our framework and the SExtractor based method. The difference between results obtained by the framework with and without the classifier is small. However, the performance of the SExtractor based method strongly depends on the classifier. Our framework is tested in a computer with Nvidia GTX 1080 Ti as GPU and two Xeon 4120 as CPU. For the same data set, it takes an average of 0.28 seconds for SExtractor to detect an image with 512 × 512 pixels, while the detection framework takes 0.21 seconds process the images of 256 × 256

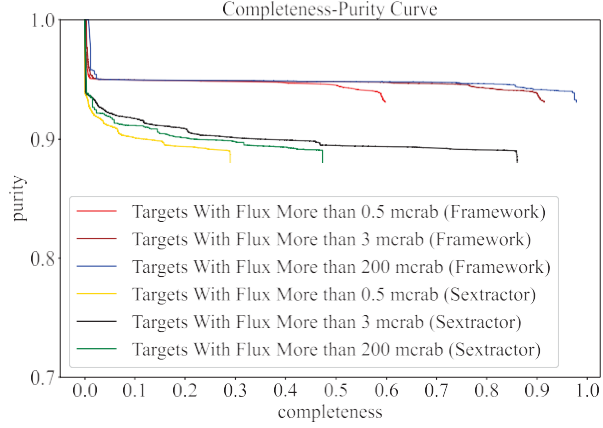

**Figure 17.** The purity-completeness curve of our detection framework and that of the SExtractor for targets with different flux ranges throughout the whole detection process

and 512 × 512 pixels. Therefore, We can find that our algorithm has almost the same speed as the SExtractor when processing images with smaller size. Since the CNN is highly parallel and could process large images if there are enough GPU memory, we could expect it has faster speed, when processing images with larger size. Based on these results, we can find that our framework has better performance in detection of transients obtained by lobster eye telescopes.

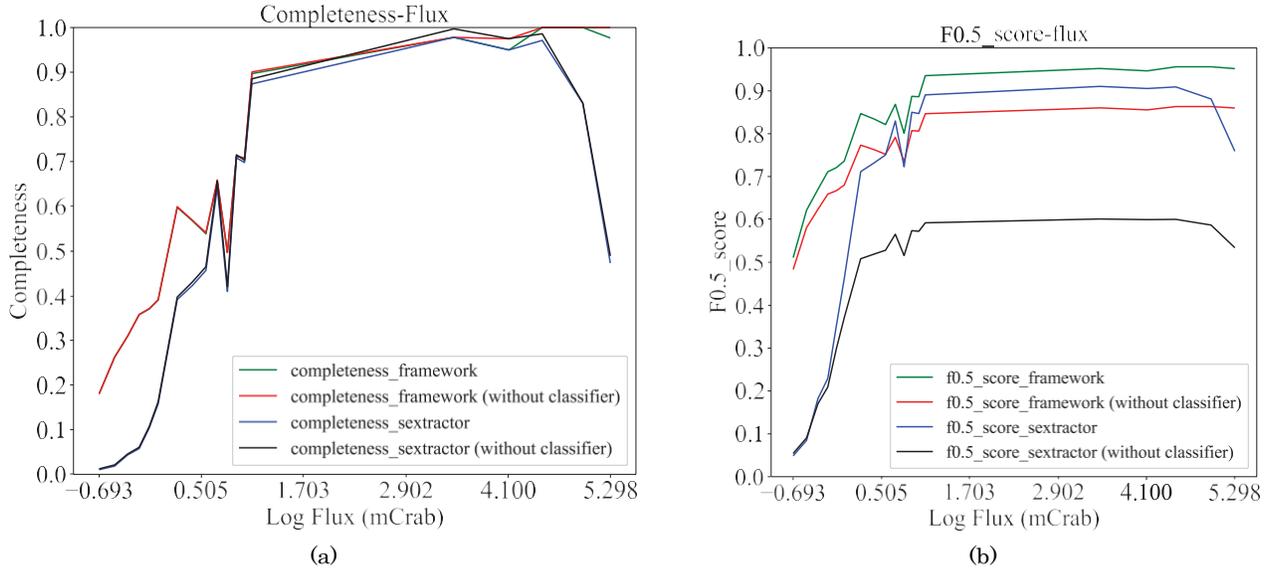

**Figure 18.** The completeness-flux and the F0.5 score-flux curves of our target detection framework and the SExtractor-based detection algorithm with and without classifier in detection targets with different flux. We bin all targets with flux more than 200 mCrab as targets with 200 mCrab. As shown in this figure, we can find that the framework is better than the SExtractor based method for most targets, especially ultra bright targets and ordinary targets.

To test the performance of our framework in detection targets with different flux, we further show the number, the completeness and average positioning error for targets with different flux in table 3. Targets with flux range from 0.5 mCrab to more than 200 mCrab are divided to 19 different bins. The interval below 1.0 mCrab is 0.1 mCrab, the interval between 1.0 mCrab and 3 mCrab is 0.25 mCrab, the interval between 3.0 mCrab and 90 mCrab is 29 mCrab and the interval between 90 mCrab and 200 mCrab is 55 mCrab. In this table, we can find that our method has a detection purity of 94%. We can also find that our target detection framework not only ensures high purity, but also has acceptable completeness (more than 50% for targets brighter than 1.25 mCrab). At the same time, the average positioning error of each flux interval is within 1.83 arcmin, which is acceptable for follow up observations. It can be seen that although the average positioning error decreases as the flux increases, the variations are small. The reason is that the neural network predicts the position of celestial objects as bounding boxes rather than the center of gravity. Our framework could provide around 0.5 pixels (1.1 arcmin) position accuracy. Higher positioning accuracy could be achieved with regression algorithms (Jia et al. 2021).

**Table 3.** Detection Results for Targets with different Flux

| Flux (mCrab) | Target Number | Completeness (Purity=0.94) | Positioning Error (arcmin) |
|---|---|---|---|
| 0.5-0.6 | 3935 | 0.1799 | 1.6584 |
| 0.6-0.7 | 2888 | 0.2621 | 1.6642 |
| 0.7-0.8 | 1766 | 0.3097 | 1.7042 |
| 0.8-0.9 | 1615 | 0.3573 | 1.7444 |
| 0.9-1.0 | 992 | 0.3700 | 1.5803 |
| 1.0-1.25 | 1347 | 0.3905 | 1.5675 |
| 1.25-1.50 | 1115 | 0.5971 | 1.6154 |
| 1.50-1.75 | 942 | 0.5669 | 1.4983 |
| 1.75-2.00 | 732 | 0.5383 | 1.4781 |
| 2.00-2.25 | 537 | 0.6555 | 1.3679 |
| 2.25-2.50 | 286 | 0.4965 | 1.3174 |
| 2.50-2.75 | 319 | 0.7116 | 1.2713 |
| 2.75-3.00 | 192 | 0.7083 | 1.4368 |
| 3.00-32.00 | 3176 | 0.9067 | 1.1223 |
| 32.00-61.00 | 320 | 0.9781 | 0.9252 |
| 61.00-90.00 | 80 | 0.9500 | 0.8788 |
| 90.00-145.0 | 70 | 1.0000 | 0.9463 |
| 145.0-200.0 | 60 | 1.0000 | 0.8599 |
| More than 200.0 | 256 | 0.9765 | 0.9167 |

To test the performance of our framework in detection of celestial objects from data obtained with different exposure time, we have generated 2D images from data with different exposure time (from 100 to 1200 seconds with interval of 100 seconds). The purity and the completeness of the detection framework are shown in figure 19. Figure 19.a shows that the purity of our detection framework could maintain over 90% for data obtained with different exposure time. To better show the completeness of our detection framework, we plot the relation between the completeness and the exposure time in figure 19.b. The red curve shows that the completeness of our algorithm increases as the exposure time increases. Meanwhile, it should be noted that according to the assumption of the distribution of targets with different flux, we would obtain more targets with low flux. Therefore, we plot the curve of completeness of our detection framework in detecting targets with flux between 0.5-3.0 mCrab, 3.0-200 mCrab and more than 200 mCrab. It can be seen that the completeness in detection of each flux interval increases as the exposure time increases. At last, we show several detection results in figure 20. As shown in this figure, we can find that most of the targets with medium brightness and even some targets that are difficult to be recognized by human eyes can be detected by our method. Overall, our algorithm is an adequate target detection framework for data obtained by the WXT onboard the EP.

## 5. CONCLUSIONS AND FUTURE WORKS

For data obtained by lobster eye telescopes, we have integrated morphological operations, the deep neural network and the random forest as a framework to detect targets from observation images. Tested with simulated data of the WXT onboard the EP, our framework can achieve high purity and moderate completeness, which enables it to be a suitable target detection algorithm. However, we find that the average positioning error of our detection framework is large, which puts forward some requirements for us to increase the positioning accuracy in the future. Besides, since we use 2D images for target detection, the information of the other two dimensions, energy and arrival time is lost. We need to further design classification algorithms based on the detection results in our future work. In the future, we would use our framework to process data obtained by the WXT onboard the EP for further scientific research.

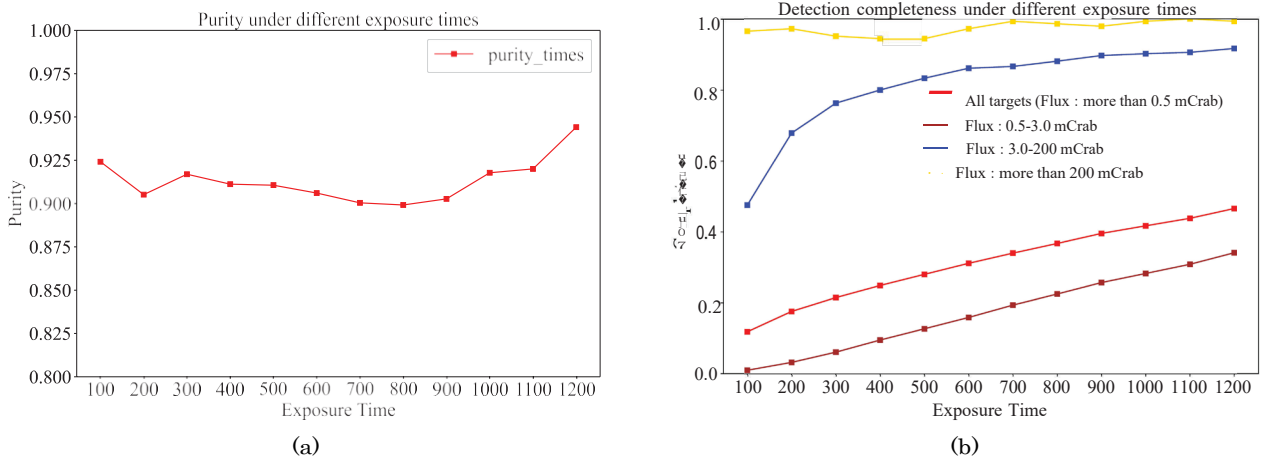

(a)                                                                      (b)

**Figure 19.** The performance of our framework in detection celestial objects from data obtained with different exposure times. Figure a shows the purity of all detection results. Figure b shows the completeness of our framework in detection of celestial objects with different flux intervals.

## ACKNOWLEDGEMENTS

Authors would like to thank the reviewer for her/his helpful suggestions. Peng Jia would like to thank Professor Weimin Yuan from National Astronomical Observatories and Professor Kaifan Ji from Yunnan Observatory who provide very helpful suggestions for this paper. This work is supported by National Natural Science Foundation of China (NSFC) with funding number of 12173027 and 12173062. We acknowledge the science research grants from the China Manned Space Project with NO. CMS-CSST-2021-A01 and CMS-CSST-2021-B12. We acknowledge the science research grants from the Square Kilometre Array (SKA) Project with NO. 2020SKA0110102. This work is also supported by the special fund for Science and Technology Innovation Teams of Shanxi Province.

## DATA AVAILABILITY

Data resources are supported by China National Astronomical Data Center (NADC) and Chinese Virtual Observatory (China-VO). This work is supported by Astronomical Big Data Joint Research Center, co-founded by National Astronomical Observatories, Chinese Academy of Sciences and Alibaba Cloud. The code and data used in this paper can be found at DOI: 10.12149/101175.

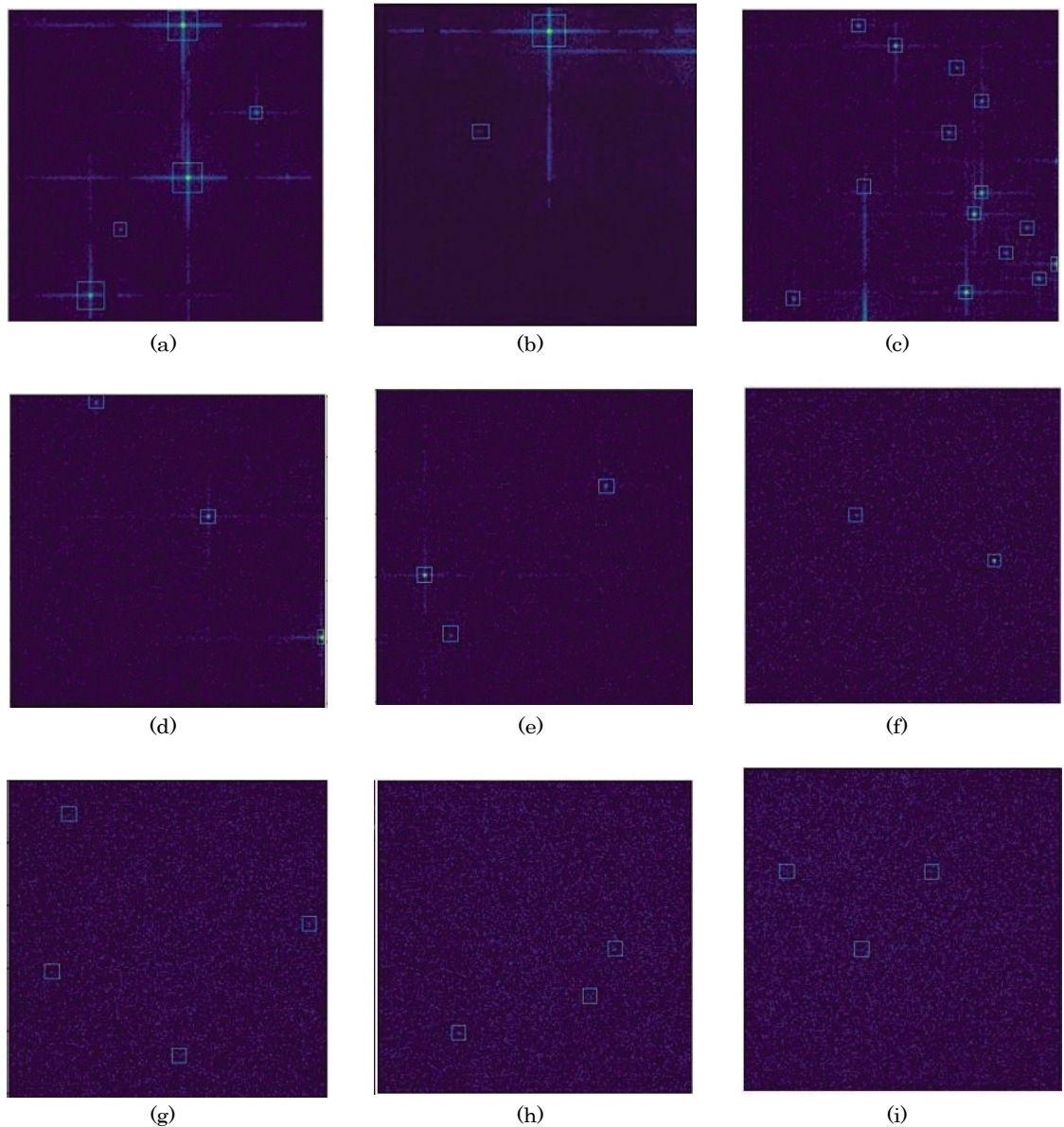

**Figure 20.** Several detection results with our algorithm. Most targets with moderate brightness (some targets are hard to seen by human eyes) could be detected our framework.